\documentstyle[aps,prb,amsmath]{revtex}
\title{%
Problem of formation of an emf in a semiconductor and its transfer
to an external circuit}
\author{%
Yu. G. Gurevich and V. B. Yurchenko}
\address{%
V. I. Lenin Polytechnic Institute, Kharkov}
\preprint{Fiz. Tekh. Poluprovodn. 25, 2109--2114 (December 1991)}
\date{(Submitted April 4, 1991; accepted for publication May 14, 1991)}
\begin{document}
\maketitle
\makeatletter
\global\@specialpagefalse
\def\@oddhead{%
Sov.~Phys.~Semicond. {\bfseries 25} (12), December 1991, p.~1268--1271
\hfill
\copyright{} 1992 American Institute of Physics}
\let\@evenhead\@oddhead

\begin{abstract}
A general description is given of a process of formation of an emf in a
medium with nonequilibrium carriers. The appearance of anomalous emfs is
predicted for several semiconductor structures. Such emfs appear as a
result of photogeneration of the majority carriers or, for example, due
to homogeneous heating of electrons and holes along the whole circuit. 
An analysis is made of the problem of determination of an emf inside a
multicomponent medium and of recording it in an external circuit.
\end{abstract}
\pacs{pacs here}
Many effects associated with the appearance of an 
electromotive force (emf) are among the topics investigated in
semiconductor physics. An emf may appear in a special
structure or in a homogeneous sample of finite 
dimensions.\cite{T62,BK77}
A theoretical description of various emfs is usually based on
models postulating some specific mechanisms of the 
appearance of an electric current and consequently different methods
of calculation of the emf are used (see, for example, Refs. 
\onlinecite{A81,RT89,EE86}). An increase in the range of the 
investigated
phenomena and the development of new semiconductor structures have
made it possible to refine the mechanisms used to account for
the observed emfs.\cite{AMPT87,GY89} However, there is as yet no 
general description of the process of formation of an emf in a 
medium containing nonequilibrium carriers. In view of the 
absence of a general treatment of the problem of how and 
because of what an emf appears, it is usual to refer to the action
of various ``external forces of nonelectrical origin,'' for
example, chemical forces. However, such a statement explains
nothing, because it does not show how, in principle, a
thermodynamic nonequilibrium gives rise to ``external forces'' and
to an electric current in a closed electrical circuit (if it does at
all) and consequently how we can calculate a possible emf in
the case of an arbitrary nonequilibrium medium. The 
conclusion that in a nonequilibrium inhomogeneous circuit the sum
of the ``contact potentials due to different carriers'' may differ
from zeros simply clouds the picture, since an electrical
potential is the same for all the carriers and the net change of
the potential. along the complete circuit is always zero, and it
is not clear what do the partial contact potentials of different
carriers represent and how to calculate them. The examples
used to illustrate such conclusions (see Refs. 
\onlinecite{T62,BK77,A81} and \onlinecite{LLP84}) usually deal only with 
those situations in which the terms
introduced can be given a very simple meaning (when the
partial contact potentials in some regions can be reduced to
differences between chemical potentials).

This means that the problem of formation of an emf
must be investigated more thoroughly. This is particularly
important in the case of a medium which contains many types
of charge carrier particularly when the energy distributions of
these carriers are far from equilibrium and the medium is 
spatially inhomogeneous. Such situations are very typical of
semiconductor structures in which electrons and holes are
readily excited by external stimuli (for example, they may be
heated by an electric field) and the appearance of any 
unexpected (and, therefore, ignored) emfs in such cases may be
of considerable importance. For example, in studies of the
transport of hot electrons in microstructures the electric field
is usually regarded as given,\cite{KH88} whereas in reality we can
expect the appearance of emfs that alter the spatial 
distribution of the field, so that its distribution must be determined in
a self-insistent manner allowing also for possible emfs.

The present paper develops a general system of concepts
on the process of formation of an emf in arbitrary conducting
structures with different nonequilibrium carriers, which would
make it possible to calculate correctly emfs of very different
origins in a great variety of situations, and to study the 
problem of how to transfer the resultant emf to an external 
electrical circuit.

We shall consider a closed circuit formed by a conducting 
material of unit cross-sectional area. We shall assume that
an electric current in this circuit is created by charge carriers
of $N$ types and each type of carrier is characterized by its
own quasi-Fermi level $F_{k}$, temperature $T_{k}$, electrical 
conductivity $\sigma_{k}$, and thermoelectric power $\alpha_{k}$ 
($k = 1, 2, \dots,N$). The partial currents of carriers in such an
electrical circuit are described by the expressions 3
\begin{equation}
j_{k} = - \sigma_{k} 
\left(
\frac{d}{dx}\tilde\varphi_{k} + \alpha_{k}\frac{d}{dx}T_{k}
\right) ,
\label{eq:gy91:jk}
\end{equation}
where $d/dx$ is the derivative with respect to the coordinate
along the circuit; 
$\tilde\varphi_{k} = F_{k}/e_{k} = \varphi + \mu_{k}/e_{k}$ is the 
electrochemical potential of carriers of the $k$th type (we shall 
consider here only the potential electric fields 
${\cal E} = -\nabla\varphi$);
$\mu_{k}$ is the chemical potential of the subsystem of carriers of the 
$k$th type. The total current $j$ is the sum of the currents $j_{k}$. 
Under steady-state conditions the total current remains constant
along the whole circuit because of the condition of continuity.
Then, the emf in such a close circuit can be described naturally 
by $E = jR$, where $R = \oint dx/\sigma$ is the total 
electrical resistance of the circuit and the conductivity is 
$\sigma = \sum_{k = 1}^{N}\sigma_{k}$.
The relationship $E = jR$ then describes Ohm's law for a
closed circuit. If we allow for Eq. (1), we find this relationship 
leads to the following general expression for the emf:
\begin{equation}
E = - \oint\sum_{k = 1}^{N}\frac{\sigma_{k}}{\sigma}
\left(
\frac{d}{dx}\tilde\varphi_{k} + \alpha_{k}\frac{d}{dx}T_{k}
\right) \, dx
\label{eq:gy91:E}
\end{equation}
where the integral is taken along the conducting circuit. 

Equation~(\ref{eq:gy91:E}) represents the most general 
description\cite{foot1} of the appearance of an emf in a closed 
electrical circuit due to the presence of carriers which are not in
thermodynamic equilibrium. Obviously, the emf appears when this integral
is not a total differentials.\cite{BK77,A79} However, the actual
conditions under which this takes place depend on the nature of the
conducting circuit and the nature of carrier nonequilibrium.

In the case of a circuit with unipolar conduction 
($N = 1$), which is in an inhomogeneous temperature field $T = T(x)$,
 the relevant conditions are described in detail in Ref. 
\onlinecite{A79}. We shall consider other possible situations.

We shall begin with the appearance of an emf in the
absence of thermal effects. It follows from Eq.~(\ref{eq:gy91:E}) that 
if
$T_{k} = {\mathrm const}$, then in a medium containing carriers of one 
kind, we have $E = 0$. This means that, irrespective of whether the
circuit with unipolar conduction is homogeneous or inhomogeneous 
and irrespective of any inhomogeneity of the generation 
of nonequilibrium carriers (of a given kind), if 
$T_{k} = \mathrm{const}$ ($k = 1$), no emf appears in the circuit (see 
also Refs. \onlinecite{BK77} and \onlinecite{A79}). It should be 
stressed that this conclusion is essentially 
related to the hypothesis that the symmetric part of the
distribution function of carriers is of the Fermi type, so that
Eq.~(\ref{eq:gy91:jk}) applies.

The situation is different in a circuit which contains
carriers of several kinds. For example, in the case of a circuit
with carriers of two kinds (usually with opposite signs), if 
$T_{k} = \mathrm{const}$ ($k = 1, 2$), we have
\begin{equation}
E = \oint \frac{\sigma_{1}}{\sigma}\frac{d}{dx}
\left(
\tilde\varphi_{2} - \tilde\varphi_{1}
\right) \, dx =
\oint \frac{\sigma_{2}}{\sigma}\frac{d}{dx}
\left(
\tilde\varphi_{1} - \tilde\varphi_{2}
\right) \, dx 
\label{eq:gy91:E12phi}
\end{equation}
or in the absence of an electrical potential $\varphi$, the 
corresponding expression is
\begin{equation}
E = \oint
\frac{\sigma_{1}}{\sigma}\frac{d}{dx}
\left(
\frac{\mu_{2}}{e_{2}} - \frac{\mu_{1}}{e_{1}}
\right) \, dx =
\oint
\frac{\sigma_{2}}{\sigma}\frac{d}{dx}
\left(
\frac{\mu_{1}}{e_{1}} - \frac{\mu_{2}}{e_{2}}
\right) \, dx
\label{eq:gy91:E12mu}
\end{equation}
In this case (when the temperature of carriers is constant) an
emf appears when, firstly, $\psi = (\mu_{2}/e_{2} - \mu_{1}/e_{1}) \ne 
\mathrm{const}$ [for example, in the case of a nondegenerate 
semiconductor
this means that the densities of nonequilibrium carriers $\delta n_{1}$, 
and $\delta n_{2}$ are not related by
\begin{equation}
\delta n_{2}(x) = 
\left[ C - \delta n_{1}(x) {n_{i}}^{2}(x) / n_{01}(x) \right]
\left[ n_{01}(x) + \delta n_{1}(x) \right] ,
\label{eq:gy91:n}
\end{equation}
where $n_{i}(x)$ is the intrinsic equilibrium density, $n_{01}(x)$ is 
the equilibrium density of carriers of the first kind, and $C$ is an
arbitrary constant] and, secondly, 
$\sigma_{1}(x)/\sigma_{2}(x) \ne \mathrm{const}$ 
(inhomogeneous medium), where $\sigma_{1}/\sigma_{2}$ 
varies along the circuit so that the integrand is no longer a total
differential.

It is obvious that these conditions for the appearance of
an emf [appropriate nonequilibrium and inhomogeneity of the
medium, and the ambipolar conduction ($N = 2$)] represent, in
particular, the familiar conditions which are necessary for the
generation of a photo-emf in solar cells (see also 
Ref.~\onlinecite{BK77}).
When these conditions (or analogous conditions in the case
when $N > 2$) are satisfied, we can expect also operation of
galvanic (``chemical'') sources of the current. If in a circuit
with such a source the value of $\psi$ varies from 
$\psi_{\mathrm{min}}$ to $\psi_{\mathrm{max}}$ and then from 
$\psi_{\mathrm{max}}$ to $\psi_{\mathrm{min}}$ in sections $a$ and $b$, 
respectively,
where $\sigma^{a} = {\sigma_{1}}^{a}$ is the electrical conductivity 
of electrons and $\sigma^{b} = {\sigma_{2}}^{b}$ is the electrical 
conductivity of ions, then the emf will be equal to its maximum
possible value $E \approx \psi_{\mathrm{max}} - \psi_{\mathrm{min}}$.

Less obvious, compared with the preceding result, is the
conclusion that follows from Eq.~(\ref{eq:gy91:E12phi}) that an emf can 
appear in a unipolar semiconductor containing several types of carriers 
of the same sign (when $T_{k} = \mathrm{const}$). Let us consider, for
example, a $p$-type semiconductor with two hole subbands
(containing light and heavy holes) where the ratio of the
mobilities depends on the coordinate. If in a certain part of
this semiconductor we create nonequilibrium holes in one of
the subbands, the diffusion of these holes gives rise to a
space charge and creates an associated electric field. This
field gives rise to an opposite drift current of both light and
heavy holes which in the open-circuit case compensates fully
the diffusion current. These processes occur on both sides of
the region with an excess hole density. If the ratio of the
mobilities of the light and heavy holes has different values on
the two sides of the region in question, then the electric fields
are also different. In this way an emf appears in the open
circuit and it is proportional to the difference between these
fields [in full agreement with Eq.~(\ref{eq:gy91:E12phi})], and when the 
circuit is closed electric current flows. This model situation can be
realize experimentally in a variable-gap semiconductor with
a coordinate-dependent ratio of the effective masses of the
light and heavy holes (for example, in Si${}_{x}$Ge${}_{1 - x}$) 
under the conditions of inhomogeneous impurity generation of
nonequilibrium holes.

In connection with this mechanism of the appearance of
an emf we should mention that the emf can appear even in a
unipolar semiconductor with one type of carrier ($k = 1$) and
with a coordinate-independent average carrier energy. In fact,
the above conclusion that such an emf cannot appear is based
on the assumption that the Einstein relationship 
$u_{k} = e_{k}D_{k}/I_{k}$
applies; here, $u_{k}$ and $D_{k}$ are the mobility and the diffusion
coefficients of carriers. if the Einstein relationship is not
obeyed (this is possible if the symmetric part of the nonequilibrium 
distribution function is not of the Fermi type), then
the ratio $u_{k}/u_{l}$ for the left- and right-hand edges of the region
of generation of nonequilibrium carriers will be different for
the same average carrier energy. In this case, as in the presence 
of two types of holes, different electric fields will appear 
on the left and right of the generation region and, therefore 
(as shown in Ref.~\onlinecite{GY89}), contrary to the generally 
accepted ideas an emf appears in a unipolar medium with a constant
average energy of carriers because nonequilibrium majority
carriers are generated.

It should be pointed out that violation of the Einstein
relationship in an inhomogeneous unipolar circuit may occur
also because of a steep drop of the average energy of carriers
in some part of the circuit when a special distribution of
heating and cooling units is adopted. This may also give rise
to an emf as a result of the mechanism discussed above. This
emf includes a contribution from a change in the thermoelectric 
power a, which is different in the regions of rise and fall
of the average carrier energy in such a circuit. These two
factors taken together are the real reason for the appearance
of the Benedicks einfl in a unipolar semiconductor.

We shall now consider the possibility of the appearance
of an emf in the presence of thermal effects and we shall do
this by returning to the temperature approximation. It should
be noted that the carrier temperature differs from the electrochemical 
potential $\tilde\varphi_{k}$ because it plays a dual role in 
Eqs.~(\ref{eq:gy91:jk})
and~(\ref{eq:gy91:E}) since it occurs in these expressions both via 
$\tilde\varphi_{k} = \tilde\varphi_{k}(T_{k})$
and directly in the form of the term 
$\alpha_{k}(dT_{k}/dx)$ (Ref.~\onlinecite{GM90}). This gives rise to
an emf even in a unipolar medium and
is responsible for the second term that is the cause of this
emf.\cite{A79} However, in media with several types of carrier there
are more opportunities for the appearance of different types
of a thermo-emf because the emf is generated not only by the
 gradients of $T_{k}$, which occur explicitly in Eqs.~(\ref{eq:gy91:jk}) 
and~(\ref{eq:gy91:E}),
but also because of the gradients of $\tilde\varphi_{k}$, which appear 
due to the dependence of $\tilde\varphi_{k}$ on $T_{k}$.

One of such unusual thermoelectric effects is, for example, 
the appearance of a thermo-emf and of a thermoelectric
current in an inhomogeneous circuit under the conditions of
spatial homogeneous heating of carriers along the whole
circuit ($T_{k} = \mathrm{const} \ne T_{0}$, where $T_{0}$ is the 
constant temperature of phonons). 

In describing this effect we shall use Eq.~(\ref{eq:gy91:E12mu}). 
Bearing in mind that the values of the chemical potentials of electrons
and holes $\mu_{n}$, and $\mu_{p}$ (measured from a constant shared 
level upward and downward, respectively) depend on the temperature 
$T_{n}$ and $T_{p}$, we find that the emf due to the heating of
carriers is
\begin{equation}
E = \frac{1}{e_{p}}\oint\frac{\sigma_{n}}{\sigma}
\frac{d}{dx}
\left(
\delta\xi_{p} + \delta\xi_{n}
\right) \, dx =
\frac{e}{e_{n}}\oint\frac{\sigma_{p}}{\sigma}
\frac{d}{dx}
\left(
\delta\xi_{p} + \delta\xi_{n}
\right) \, dx
\label{eq:gy91:Exi}
\end{equation}
where 
\[
\delta\xi_{k} = \varepsilon_{k} - \varepsilon_{k0} = 
(T_{k}/T_{0} - 1)\xi_{k0} + T_{k}[\ln(n_{k}/n_{k0}) -
 3/2\ln(T_{k}/T_{0})];
\]
\[
\xi_{k} = \xi_{k}(T_{k}, n_{k}) = T_{k}\ln[n_{k}N_{k}(T_{k})];
\]
\[
\xi_{k0} = \xi_{k}(T_{0}, n_{k0});
\] 
$n_{k0}$ and $n_{k}$ are the equilibrium and
nonequilibrium carrier densities; $N_{k}(T_{k})$ is the effective 
density
of states in the relevant band (it is assumed that the investigated 
semiconductor is nondegenerate). If the density of the
majority and minority carriers does not change during heating, 
it follows from Eq.~(\ref{eq:gy91:Exi}) that the emf is described by the
expression
\begin{equation}
E = - \frac{\vartheta_{p}}{e_{p}} \oint\frac{\sigma_{n}}{\sigma_{n} + 
\sigma_{p}} 
\left[
\frac{dE_{g}}{dx} + 
\left(
1 - \frac{\vartheta_{n}}{\vartheta_{p}}
\right)
\frac{d\xi_{n0}}{dx}
\right] \, dx
\label{eq:gy91:Etheta}
\end{equation}
where $\vartheta_{k} = (T_{k} - T_{0})/T_{0}$ and where $E_{g}$ is the 
band
gap of the semiconductor. It is clear from Eq.~(\ref{eq:gy91:Etheta}) 
that
in a closed circuit with an inhomogeneous doping and particularly in one
with a variable band gap an emf may indeed appear when the
heating of carriers is homogeneous along the whole circuit. If
$T_{n} = T_{p} \ne T_{0}$, this is possible only in a variable-gap 
circuit
[it is understood that naturally this requires a suitable inhomogeneous 
doping so that the value of $\sigma_{n}/(\sigma_{n} + \sigma_{p})$ 
varies continuously]. Even if $E_{g} =\mathrm{const}$, an emf may 
appear if the
heating of electrons and holes is different ($T_{n} \ne T_{p}$; in 
particular, $T_{n} \ne T_{p} = T_{0}$) and, consequently, the carrier 
mobility depends on the coordinate [the coordinate dependence of
just the carrier density is insufficient, since then the quantity
$\sigma_{n}/(\sigma_{n} + \sigma_{p})d\xi_{n0}$ is not a 
total differential].

In spite of the very special nature of the situation discussed 
here, the possibility of the appearance of such a thermal-emf 
is of fundamental important because the conditions
needed for the generation of anomalous emf s may occur if
not throughout the circuit then at least in some parts of it.
For example, this thermo-emf is closely related to the familiar 
hot-carrier thermo-emf across a $p$-$n$ junction.\cite{BPAD75}

In fact, if we allow for the continuity of the quasi-Fermi
levels of electrons and holes across a $p$-$n$ junction, then as
emf which appears in the case of homogeneous heating of
carriers in the vicinity of a symmetric junction when the
circuit is open and the temperatures $T_{n}$ and $T_{p}$ are identical,
can be described by the following expression which is deduced 
from Eq.~(\ref{eq:gy91:Exi}):
\begin{equation}
E_{pn} = 
\left(
\delta\xi_{p}^{(n)} - \delta\xi_{p}^{(p)}
\right) / \bar{e}_{p} ,
\label{eq:gy91:Epn}
\end{equation}
which in the $n_{k} = n_{k0}$ ($k = n, p$) case gives the familiar
result\cite{BPAD75}
\begin{equation}
E_{pn} = U_{pn} \left(T_{p,n} - T_{0} \right)/T_{0} , 
\label{eq:gy91:Eu}
\end{equation}
where $U_{pn} = [\xi_{p}^{(n)}(T_{0}) - \xi_{p}^{(p)}(T_{0})]/e_{p}$ is 
the equilibrium
contact potential across a junction (the upper indices identify
the $p$- and, $n$-type regions of the junction). However, it should
be noted that an important condition in the derivation of 
Eq.~(\ref{eq:gy91:Eu}) is
the constancy of the densities of the majority and minority 
carriers during heating. However, the densities of
carriers of one or the other kind can in fact vary with heating, 
so that the value of the emf may differ from that given
by Eq.~(\ref{eq:gy91:Eu}). In particular, if the 
generation-recombination
equilibrium between the energy bands is controlled by direct
band-band transitions, which is typical of semiconductors
with a sufficiently narrow band gap, the heating of carriers
causes their densities to rise in the same way as if they were
heated together with the lattice. Then, if $T_{n} = T_{p} \ne T_{0}$, 
the positions of the quasi-Fermi levels of electrons and holes in
the band gap coincide (as in the $T_{n} = T_{p} = T_{0}$ case) and they
shift on increase in the difference ($T_{n,p} - T_{0}$). Consequently,
the emf $E_{p,n}$ described by Eq.~(\ref{eq:gy91:Epn}) vanishes. If the 
carrier
temperatures at the external contacts are then equal to $T_{0}$, we
have the usual bulk thermo-emf $E_{T}$, which appears also when
carriers are heated together with the phonons. The latter
thermo-emf is much less than $E_{pn}$ in Eq.~(\ref{eq:gy91:Eu}) and has 
the opposite sign.

We shall now go back to our general case of a closed
circuit in which ad cmf of arbitrary physical nature is generated. 
It is worth noting that such an electrical circuit can be
divided into a region where an emf is generated and a region
representing an external load only if the circuit has a section
where $n_{k} = n_{k0}$ and $T_{k} = T_{0}$ for all Idnds of carriers. 
It is this section that plays the role of an external load. if there is
no such section, then in any selected part of the closed circuit
the concept of the emf formed in this section becomes ambiguous 
and this is true also of the voltage drop across this
section. For example, in the case of an ambipolar semiconductor 
when $T_{n} = T_{p} = T_{0} = \mathrm{const}$, we have the following
obvious system of equations
\begin{multline}
jr \equiv j \int_{a}^{b}\sigma^{-1}dx = \int_{a}^{b}
\left( 
\frac{\sigma_{1}}{\sigma}\frac{d\tilde\varphi_{1}}{dx} +
\frac{\sigma_{2}}{\sigma}\frac{d\tilde\varphi_{2}}{dx}
\right) \, dx \\
= - \int_{a}^{b}d\tilde\varphi_{1} + 
\int_{a}^{b}\frac{\sigma_{2}}{\sigma} 
\frac{d}{dx}\left(\tilde\varphi_{1} - \tilde\varphi_{2}\right) \, dx\\
= \Delta\tilde\varphi_{1} + E_{1} = - \int_{a}^{b}d\tilde\varphi_{2} +
\int_{a}^{b}\frac{\sigma_{1}}{\sigma}\frac{d}{dx}\left(\tilde\varphi_{2}
- \tilde\varphi_{1}\right) \, dx = \Delta\tilde\varphi_{2} + E_{2} .
\label{eq:gy91:jrT}
\end{multline}
If by a section of a circuit we understand the whole closed
contour ($r = R$), then $E_{1} = E_{2} = E$ [compare with 
Eq.~(\ref{eq:gy91:E12phi})].
It is clear from Eq.~(\ref{eq:gy91:jrT}) that if at the points $a$ and 
$b$ there is
no carrier equilibrium ($\tilde\varphi_{1}\ne\tilde\varphi_{2}$), then 
in general
we have $\Delta\tilde\varphi_{1}\ne\Delta\tilde\varphi_{2}$ and, 
consequently, $E_{1} \ne E_{2}$.
However, if in spite of nonequilibrium we have 
$\Delta\tilde\varphi_{1} = \Delta\tilde\varphi_{2}$ then
it would seem that the separation of the quantity $jr$ into a voltage 
drop $\Delta\tilde\varphi_{ab}$ and an emf $E_{ab}$ is unambiguous, the 
readings of a voltmeter connected between the points $a$ and $b$ do not 
give $\Delta\tilde\varphi_{ab}$. This is due to the fact that a 
separate
emf appears in this case in the voltmeter circuit and this emf is due to
nonequilibrium conditions. An ideal voltmeter is a device
which does not alter the current in the measuring circuit
(which means that the resistance of the voltmeter should be
infinite), does not influence the carrier nonequilibrium, and
does not develop its own emf. From all this it follows that
the concept of a voltage drop can be introduced only for parts
of a circuit between the points with equilibrium carriers and
the voltmeter must be connected to these points. The voltage
drop should then be the quantity $\Delta\tilde\varphi_{ab} \equiv 
\Delta\tilde\varphi_{1} \equiv \Delta\tilde\varphi_{2}$,
which is measured directly by the voltmeter. This quantity is equal
 to the drop of the electrochemical potential of carriers between 
the points $a$ and $b$, which is the same for all the carrier
 subsystems (irrespective of whether the electrochemical potentials
 of different carriers are the same within the investigated
 region). In the case of an electrical potential, its drop
$\Delta\varphi_{ab}$ differs from that measured by the voltmeter 
$\Delta\tilde\varphi_{ab}$
by an amount $\Delta\mu_{ab}$, which is not equal to zero for an 
inhomogeneous circuit.

It follows from the above discussion that in the case of a
unipolar semiconductor with the Fermi-type symmetric part
of the distribution function, when the emf is related only to
an inhomogeneity of the temperature distribution, the voltage
drop should strictly speaking by determined between points at
the same temperature. Clearly, if the intrinsic thermo-emf of
a voltmeter vanishes, this voltmeter gives a reading of 
$\Delta\tilde\varphi_{ab}$
between any points and in the more usual general situation
this can naturally be called the voltage drop. It should be
pointed out that in the traditional approach to the definition of
the thermo-emf it is $\Delta\tilde\varphi$ which is implied and not 
$\Delta\varphi$.
This is attributed in Ref.~\onlinecite{A81} to the fact that the 
quantity $\Delta\varphi$ at
the contacts has a discontinuity, whereas $\Delta\tilde\varphi$ is 
continuous.
However, in fact $\Delta\tilde\varphi$ may also have a discontinuity 
(this is true
if the conductivity of the contact itself is finite). In this case the
jumps $\Delta\varphi$ may be associated with their own emfs influencing
the readings of the instrument (as observed in the case of a
hot-carrier $p$-$n$ junction if the contact is understood to be the
whole $p$-$n$ junction region where the heating takes place).
Consequently, in accordance with the conclusions reached in
the present paper, we can find the emf if we determined 
$\Delta\tilde\varphi$
at the ends of a region which includes all the discontinuities
of the electrochemical potentials of carriers of each kind.

All this is valid not only in the case of finite but also in
the case of infinitesimally short sections of the circuit. Therefore,
if an external voltage is applied to some part of a circuit
and inside this part there are nonequilibrium carriers capable
of creating an emf (this nonequilibrium state may be induced,
in particular, by the applied voltage itself), then the electric
field at the internal points in this part cannot be separated
unambiguously into the purely ``external'' field and the ``internal''
(``nonequilibrium built-in'') field, which is associated
with the generated emf. Similarly, a change in the electrical
potential inside the medium on appearance of a nonequilibrium
creating an emf cannot be interpreted as the emf itself
(compare with Ref.~\onlinecite{RT89}), but outside the medium the emf 
is an indeterminate and directly measurable quantity.

An analysis of the process of formation of an emf given
above thus provides a clear physical picture of the possible
mechanisms and the conditions for the appearance of an emf
of any nature in arbitrary electrical circuits with nonequilibrium
carriers and it provides definite procedures for the calculation
of such emf's.

Translated by A. Tybulewicz


\begin{references}
\bibitem{T62}
J. Tauc,
{\itshape Photo- and Thermoelectric Effects in Semiconductors, \/}
Pergamon Press, Oxford (1962). 
\bibitem{BK77}
V. L. Bonch-Bruevich and S. G. Kalashnikov, 
{\itshape Physics of Semiconductors\/} [in Russian], 
Moscow (1977).
\bibitem{A81}
A. I. Anselm, 
{\itshape Introduction to Semiconductor Theory,\/} 
Mir, Moscow; Prentice-Hall, Englewood Cliffs, NJ (1981).
\bibitem{RT89}
B. I. Reznikov and G. V. Tsarenkov, 
Fiz. Tekh. Poluprovodn. {\bfseries 23}, 1235 (1989) 
[Sov. Phys. Semicond. {\bfseries 23}, 771 (1989)].
\bibitem{EE86}
A. V. Efanov and M. V. Entin, 
Fiz. Tekh. Poluprovodn. {\bfseries 20}, 20 (1986)
[Sov. Phys. Semicond. {\bfseries 20}, 11 (1986)].
\bibitem{AMPT87}
V. L. Al'perovich, S. P. Moshchenko, A. G. Paulish, and 
A. S. Terekhov, 
Fiz. Tekh. Poluprovodn. {\bfseries 21}, 324 (1987) 
[Sov. Phys. Semicond. {\bfseries 21}, 195 (1987)].
\bibitem{GY89}
Yu. G. Gurevich and V. B. Yurchenko, 
Solid State Commun. {\bfseries 72}, 1057 (1989).
\bibitem{LLP84}
L. D. Landau, E. M. Lifshitz, and L. P. Pitaevskii, 
{\itshape Electrodynamics of Continuous Media,\/} 
2nd ed., Pergamon Press, Oxford (1984).
\bibitem{KH88}
K. Kim and K. Hess, 
Solid State Electron. {\bfseries 31}, 877 (1988).
\bibitem{foot1}
Naturally, if $T_{k}$ and $F_{k}$ can be introduced at all (for details
see later). 
\bibitem{A79}
L. I. Anatychuk, 
{\itshape Thermoelements and Thermoelectric Devices 
(Hand book)\/} [in Russian], Kiev (1979).
\bibitem{BZM75}
V. L. Bonch-Bruevich, I. P. Zvyagin, and A. G. Mironov, 
{\itshape Domain Electrical Instabilities its Semiconductors,\/}
Consultants Bureau, New York (1975).
\bibitem{GM90}
Yu. G. Gurevich and O. L. Mashkevich, 
Fiz. Tekh. Poluprovodn. {\bfseries 24}, 1327 (1990) 
[Sov. Phys. Semicond. {\bfseries 24}, 835 (1990)].
\bibitem{BPAD75}
A. I. Beinger, L. G. Paritskii, E. A. Akopyan, and G. Dadamirzaev,
Fiz. Tekh. Poluprovodn. {\bfseries 9}, 216 (1975) 
[Sov. Phys. Semicond. {\bfseries 9}, 144 (1975)].
\end{references}
\end{document}